# Characterizing AI Agents for Alignment and Governance

Atoosa Kasirzadeh[1] & Iason Gabriel[2]

**Abstract**. The creation of effective governance mechanisms for AI agents requires a deeper understanding of their core properties and how these properties relate to questions surrounding the deployment and operation of agents in the world. This paper provides a characterization of AI agents that focuses on four dimensions: autonomy, efficacy, goal complexity, and generality. We propose different gradations for each dimension, and argue that each dimension raises unique questions about the design, operation, and governance of these systems. Moreover, we draw upon this framework to construct "agentic profiles" for different kinds of AI agents. These profiles help to illuminate cross-cutting technical and non-technical governance challenges posed by different classes of AI agents, ranging from narrow task-specific assistants to highly autonomous general-purpose systems. By mapping out key axes of variation and continuity, this framework provides developers, policymakers, and members of the public with the opportunity to develop governance approaches that better align with collective societal goals.

## 1. Introduction

AI researchers have long been interested in artificial intelligence (AI) agents, ranging from reinforcement learning (RL) agents to autonomous vehicles (Feigenbaum, 1977; Russell and Norvig, 1995; Sutton and Barto, 2020; Wooldridge, 2000). However, recent breakthroughs have led to the development of a new class of AI agents based on powerful foundation models – which are then supplemented with scaffolding for advanced reasoning capabilities, memory, and tool use (Sumers et al., 2023). Building on this architecture, we will likely see a large number of novel AI agents deployed across a range of real-world domains in the near future.[3] Such agents could take a variety of different forms (Casper et al., 2025), including advanced AI assistants (e.g. Google DeepMind's Astra), digital companions (e.g. Replika), AI researchers or tutors (e.g. Snorkl), AI workers, and new types of autonomous robots.

To better understand this trend and render it tractable, we review various definitions of AI agents that are currently influential, critically examine their component parts, and develop a framework that tracks key dimensions of AI agency, with a particular focus on the implications of these dimensions for technical and non-technical AI governance. We characterise AI agents as systems that have the ability to perform increasingly complex and impactful goal-directed action across multiple domains, with limited external control. In essence, our focus is on a large class of artificial systems that are able to independently pursue a wide range of goals and tasks, thereby exerting causal influence on the world.

---

[1] Carnegie Mellon University; atoosa@cmu.edu
[2] Google DeepMind, London; iason@google.com
[3] Since 2023, major technology and business organizations have accelerated their development of foundation model-based "AI agents". To name just a few examples: McKinsey Quarterly published a report titled "Why agents are the next frontier of generative AI" (Yee et al., 2024); Salesforce (2024) unveiled a suite of agents designed to handle customer tasks across various functions, setting an ambitious goal to deploy "one billion agents" by the end of 2025; and Microsoft advanced this trend by announcing new autonomous agent capabilities (Spataro, 2024) in Copilot Studio and its Dynamics 365 suite, aiming to transform business processes with AI. In addition, Anthropic (2024a) has introduced computer interaction capabilities powered by Claude 3.5, enabling the AI to view screenshots and perform basic computer actions like moving cursors and clicking when using compatible software. This allows Claude to interact with computer interfaces in ways that parallel human computer use.

The broad deployment of AI agents may have a profound effect on existing social, economic and political practices (Gabriel et al., 2024). For example, AI agents may disrupt labour markets by autonomously performing cognitive tasks that previously required human workers, such as analyzing legal documents or writing software code. They could also fundamentally reshape personal interactions with AI, becoming an ever-present interlocutor, executor, and source of advice for humans, who may form parasocial relationships with them (Kirk et al., 2025).

In light of this, we need to understand and categorize different kinds of AI agent, explore their implications, assess their governance requirements, and gear efforts towards responsible deployment. In particular, we need to guard against both immediate individual risks and systemic challenges that might arise from the deployment and use of these systems (Kasirzadeh, 2025). At the individual level, this includes preventing accidents and malicious use (Chan et al., 2023; Uuk et al., 2024). At the collective level, this means understanding and navigating the complex dynamics that emerge when multiple human and AI agents interact with one another, including coordination failures and harmful competition or collusion between agents (Hammond et al., 2025). Practically speaking, governance structures also need to address the question of how to assign liability and responsibility for AI agent actions (Kolt, 2025), how to implement monitoring and auditing procedures for different domains of AI agent operation, how to create permissible action spaces and authentication protocols for these agents, and how to build infrastructure for agent-to-agent interaction (Chan, 2025) including identification requirements (Chan, 2024).

Fortunately, we can draw on a number of existing frameworks when it comes to governing AI agents. For example, autonomous vehicle regulation has typically used one schema for measuring the level of autonomy evidenced by AI systems in this domain, providing guidance that is tailored to the level of autonomy a vehicle demonstrates (Society of Automotive Engineers, 2021). Other approaches, such as the EU AI Act (Council of the European Union, 2024) and the National Institute of Science and Technology's AI Risk Management Framework (Tabassi, 2023), aim to classify the overall levels of risk attached to an AI system and to employ safeguards that are proportionate to this risk. However, both approaches could benefit from further engagement with the novel properties of AI agents. By mapping and understanding different dimensions of AI agency, we aim to show that new cross-cutting governance issues come to light.

This paper develops a conceptual framework for mapping characteristics of AI agency along four dimensions, develops a suggested set of categories that scale for each dimension, and discusses the governance implications of these dimensions. Section 2 explores prevalent conceptions of AI agency found in computer science and identifies core constitutive properties that unify these accounts. The subsequent sections (3–6) then provide a more detailed specification of each property, justification for this interpretation, and a set of gradations that illustrate what higher or lower levels of property-endowment look like. Section 7 uses the framework to cast light on the properties of four actually existing agentic AI systems, developing "agentic profiles" for AlphaGo, ChatGPT-3.5, Claude 3.5 with tool use, and Waymo. Finally, Section 8 discusses the governance implications that arise from a better understanding of different kinds of AI agency and identifies key avenues for future research.

## 2. Definitions of AI Agency

Efforts to understand and characterize "agency" have a long-standing history. The philosophical[4] and legal[5] foundations of agency have been explored extensively, with different theories aiming to explain the presence – or absence – of agency in systems, organisms, and organizations. Researchers have sought to analyze different types of agency, including biological,[6] epistemic,[7] group-based,[8] and shared agency,[9] which extends to encompass decisions reached jointly by multiple actors. These conceptions have, in various ways, influenced the understanding of agency found in computer science – one that has, in turn, played a foundational role in the development of modern AI systems.

---

[4] Contemporary philosophy offers three main views for understanding agency. First, the *intentional view* roots agency in purposeful action. It maintains that being an agent necessarily involves acting with intention, based on mental states like beliefs and desires (Goldman, 1970; Davidson, 1971; Dretske, 1988; Dung, 2024). For example, when a person deliberately reaches for a coffee cup, their desire for coffee and belief about the cup's contents cause the intentional action. Second, and in contrast, the *non-intentional view* (Ginet, 1990; O'Connor, 2002; Lowe, 2008) argues that authentic agency can emerge without prior intentional planning. Examples include spontaneously jumping up to dance when hearing music or reflexively catching a falling object. This view rejects the necessity of causal relations between mental states and actions. Third, the *hierarchical view,* distinguishes between types of agency by emphasizing the capacity for meta-cognition (Frankfurt, 1971; Taylor, 1977): while a dog might simply eat when hungry (basic desire), humans can reflect on their eating habits ("Why do I stress eat?"), form attitudes about their attitudes ("I shouldn't let anxiety drive my eating"), and create self-governing policies ("I'll practice mindful eating at dinner"). The characterization of AI agency developed in this article differs from these accounts insofar as it does not require that agents have mental states that are analogous to those of human beings. For example, an autonomous robot may navigate a room successfully and conduct complex goal-directed activities in the space without having beliefs *about the room* that resemble those of a human agent.

[5] The legal and economic literature provides another crucial lens for understanding AI agency (Kolt, 2025). For example, the principal–agent paradigm specifies that one entity (an agent) can act on behalf of another (principal) under certain conditions, creating chains of authority and responsibility, as is the case for corporate directors' fiduciary duties to shareholders or attorneys' obligations to clients (Grossman et al., 1992). Similarly, the common law doctrine of agency relationships offers a valuable framework for analyzing AI agents. Agency law principles concerning information disclosure, scope of authority, fiduciary duties of loyalty, and rules governing delegation to subagents are particularly relevant.

[6] Biological agency examines how living organisms – from single cells to complex organisms – act purposefully in their environments. See Kauffman and Clayton(2006), Bouvard (2009), and Meincke (2018).

[7] Epistemic agency focuses on how agents gather, evaluate, and act on knowledge, from scientists conducting research to students managing their learning. See Damşa et al. (2010), Elgin (2013), and Sosa (2013).

[8] Group agency explores how collectives – from ant colonies to corporate boards – can act as coherent agents despite being composed of individual actors. See Nickel (1997), List and Pettit, P. (2011), and Tollefsen (2015).

[9] Shared agency examines how individuals coordinate to achieve common goals, as seen in everything from two people moving furniture to massive orchestra performances. See Bratman (2014), Shapiro (2014), Le Besnerais et al. (2024).

For example, the notion that agency requires the capacity for rational deliberation and intentional action based on reason has been particularly influential, informing work by AI researchers such as Russell (1997) and Rao & Wooldridge (1999) who sought to build AI systems that could reason about goals and actions in response to environmental feedback. In fact, the first edition of Russell & Norvig's leading textbook, *Artificial Intelligence: A Modern Approach*, originally published in 1995, prominently featured the phrase "intelligent agent" on its cover, highlighting the early recognition of agency as a fundamental concept for AI.[10]

In recent years, the idea of AI agency has evolved further along several different developmental paths. RL has continued to focus on autonomous AI agents that learn optimal behaviours through environmental interaction and feedback (Sutton and Barto, 1998); these agents have achieved remarkable success, particularly in the context of game environments (Silver et al., 2017; Brown and Sandholm, 2019; Meta Fundamental AI Research Team (FAIR) et al., 2022).[11] Agent-based modelling and simulation-based approaches have emphasized the importance of social interaction and emergence in the context of building more capable artificial systems (Wooldridge, 1999). Recent work has also explored agents that are capable of limited forms of meta-learning, or learning-to-learn, and self-modification (Lake et al., 2017, Park et al., 2023). Finally, progress has been made in the domain of reasoning, with Zelikman et al. (2022) demonstrating that language models can improve this capability through iterative self-correction and fine-tuning on successfully generated reasoning paths. The incorporation of reasoning protocols sits at the heart of many recent advances achieved by models that can "think", such as Claude 3.5 (Anthropic, 2024b), Gemini 2.5 (Kavukcuoglu, 2025), and OpenAI's o1 model (Jaech, 2024).

---

[10] Franklin & Graesser's (1997) groundbreaking work established fundamental distinctions between AI agents and traditional software programs, paving the way for key debates about the nature of agency in AI research. Their taxonomy of autonomous agents provided an early account of core characteristics that distinguish reactive programs from genuine artificial agents – such as independent goal pursuit, environmental responsiveness, and temporal continuity. This theoretical foundation coincided with research on social agency in AI systems. For instance, Elliott & Brzezinski (1998) investigated how autonomous agents could function in social environments, developing synthetic personas capable of social interaction. Parallel work by Nass and colleagues (1994) established the "computers as social actors" paradigm, demonstrating how humans naturally attribute social characteristics to computational agents and interact with them using social rules and expectations. This research, rooted in human factors and human–computer interaction studies, showed how people readily engage with artificial agents as social entities, even while consciously aware of their non-human nature. These insights influenced the development of educational agents (Payr, 2003), where artificial tutors and learning companions leverage social agency to enhance educational outcomes.

[11] The RL paradigm provides a framework for understanding how agents learn through interaction to achieve goals (Sutton and Barto, 1998). In this paradigm, the learner –which is the agent – interacts with its environment through a cycle of observation, action, and reward. They define "the reinforcement learning problem" as "the problem of learning from interaction what to do—how to map situations to actions—so as to maximize a numerical reward signal" (Sutton & Barto, 1998, p.1). At its core, RL formalizes goal-directed learning through an objective function – the mathematical expression of what the agent aims to achieve, typically represented as the maximization of cumulative rewards over time. An RL agent learns by interacting with an environment. It uses sensors to perceive the environment's state and actuators to take actions within a defined state-action space. The environment provides feedback, allowing the agent to learn how to achieve its goals. The complexity of the state–action space fundamentally determines the difficulty of the learning problem. In some cases, the agent's objective function is not explicitly defined but must be inferred through techniques like inverse RL, which reconstructs reward functions from observed behaviour (Ng and Russell, 2000). This approach has proven particularly valuable in understanding and replicating complex behaviours where goals may be implicit rather than explicitly stated, allowing for more nuanced modelling of agent objectives and decision-making processes.

Yet, it is worth asking: What unifies these different accounts of AI agency? Along what dimensions – and across what features – do these different conceptions come together or draw apart? The table below facilitates comparison across seven key accounts found in AI and machine learning literature.

| | |
|---|---|
| **D.1: Russell & Norvig (1995)** | "An agent is anything that can be viewed as perceiving its environment through sensors and acting upon that environment through effectors." |
| **D.2: Franklin & Graesser (1997)** | "An autonomous agent is a system situated within and [is] part of an environment that senses that environment and acts on it, over time, in pursuit of its own agenda and so as to affect what it senses in the future." |
| **D.3: Wooldridge (1999)** | "Agents are simply computer systems that are capable of autonomous action in some environment in order to meet their design objectives. An agent will typically sense its environment (by physical sensors in the case of agents situated in part of the real world, or by software sensors in the case of software agents), and will have available a repertoire of actions that can be executed to modify the environment, which may appear to respond non-deterministically to the execution of these actions." |
| **D.4: Park et al. (2023)** | "Generative agents [are] computational software agents that simulate believable human behavior. Generative agents wake up, cook breakfast, and head to work; artists paint, while authors write; they form opinions, notice each other, and initiate conversations; they remember and reflect on days past as they plan the next day." |
| **D.5: Huang et al., (2024)** | "Autonomous agents have been recognized as intelligent entities capable of accomplishing specific tasks, via perceiving the environment, planning, and executing actions." |
| **D.6: Masterman et al. (2024)** | "AI agents are language model-powered entities able to plan and take actions to execute goals over multiple iterations. AI agent architectures are either comprised of a single agent or multiple agents working together to solve a problem… [T]he research community has experimented with building autonomous agent-based systems." |

| D.7: Shavit et al. (2023) | "Agentic AI systems are characterized by the ability to take actions which consistently contribute towards achieving goals over an extended period of time, without their behavior having been specified in advance." |

**Table 1. Seven key definitions of AI agents from computer science**

In the following sections, we identify the commonalities, distinctions, strengths, and weaknesses of these accounts, unpacking what we take to be the four core constitutive properties of AI agency – autonomy, efficacy, goal complexity, and generality – and using these properties to construct "agentic profiles" for different kinds of AI agents: Alpha Go, a Waymo autonomous vehicle, ChatGPT 3.5 (stand-alone chat completion), and Claude Sonnet 3.5 (with tool use).

## 3. Autonomy

Autonomy is a key property of all definitions of AI agents. Across various definitions, researchers repeatedly foreground the autonomous "action" (D.1, D.3, D.7) of AI systems. Other accounts focus on "autonomous agents" (D.2, D.5, D.6) or autonomous behaviours, such as forming opinions without supervision (D.4).

From a philosophical standpoint, autonomy can be understood in many ways, including as negative freedom – understood as freedom from certain external constraints – and positive freedom – understood as the capacity to pursue goals without assistance (Berlin, 1969). Autonomy is also sometimes connected to the range of actions, or "degrees of freedom", which an agent is in a position to take (Dennett, 2003). In the context of AI agents, agency is best understood in terms of the capacity to perform actions *without external direction or control*. This characterization captures what makes AI agents distinctive: they can independently determine and execute sequences of action (directed toward specific goals) without requiring step-by-step guidance. In most cases, the relevant form of external direction or control comes from a principal comprised by a single human or set of humans, although in some cases it could come from another AI system or control mechanism.

Despite this common concern with autonomy across various definitions of AI agents, the field as a whole lacks a standardized way of classifying or measuring the degrees of autonomy an AI agent possesses. Levels of autonomy matter in the context of AI agent governance, as they help to determine where and when different kinds of oversight mechanism are needed – for example, distinguishing between AI agents that merely execute predefined workflows (which may require minimal oversight) versus those that act independently (which may demand more rigorous safety verification and monitoring protocols). Without such gradations, governance frameworks risk applying identical standards to qualitatively different AI agents, creating either unnecessary barriers or dangerous oversight gaps.

Fortunately, the conception of autonomy has been modelled for autonomous vehicles (Vagia et al., 2016; Society of Automotive Engineers, 2021) by a categorical scale.[12] We believe that

---

[12] The Society of Automotive Engineers has established a framework for classifying autonomous vehicles that defines six levels of driving automation: Level 0 (No Driving Automation), where the human driver performs all driving tasks; Level 1 (Driver Assistance), where the vehicle has a single

this scale can be helpfully adapted to characterize the autonomy of AI agents in general. Relevant gradations of AI autonomy are shown in Table 2:

| A.0: No autonomy | The AI system is *entirely dependent* upon the principal for its ability to act and can only act in the manner the principal dictates. |
|---|---|
| A.1: Restricted autonomy | The AI system can conduct a single automated task. The other tasks always take place under the principal's *direct oversight*. |
| A.2: Partial autonomy | The AI system can conduct a range of automated tasks. The principal must remain engaged and be ready to take control at any time. |
| A.3: Intermediate autonomy | The AI system can perform the majority of tasks independently, though it still relies upon *input* from the principal for critical determinations. |
| A.4: High autonomy | The AI system can independently perform all tasks *in certain circumstances*, though oversight is maintained by the principal when those circumstances are not met (in the event of aberrant behaviour). |
| A.5: Full autonomy | The AI system is able to perform all tasks *without oversight or control*. |

**Table 2. Levels of autonomy for AI agents**

We will shortly say more about the relevance of autonomy for AI governance challenges. At this juncture, however, we want to draw attention to two things. First, AI autonomy is especially important in the context of AI safety (Chan et al., 2023; Anwar et al., 2024; Hammond et al., 2025). As AI systems become increasingly autonomous, they have the potential to unlock significant value – reducing the need for human effort and labour in order to achieve goals. However, this benefit is achieved at a cost: higher levels of autonomy can only be achieved via a reduction in the frequency and strength of external direction or oversight. Second, full autonomy (A.5) for AI systems is not a desirable goal – given that it entails the loss of the principal's control – unless the efficacy of a system is wholly

---

automated system for driver assistance; Level 2 (Partial Driving Automation), where the vehicle has combined automated functions but the driver must remain engaged; Level 3 (Conditional Driving Automation), where the vehicle can handle all aspects of driving under certain conditions but the driver must be ready to take control; Level 4 (High Driving Automation), where the vehicle performs all driving functions under specific conditions with no driver intervention needed; and Level 5 (Full Driving Automation), where the vehicle performs all driving functions under all conditions with no driver needed.

contained, its capabilities are robustly aligned, and its capabilities significantly limited (*see* 4. Efficacy).[13]

## 4. Efficacy

A second key feature of agents is their ability to *interact with and have a causal impact upon the environment*. The property of efficacy is found in each definition of AI agency, which describe agents "sens[ing] the environment and act[ing] on it" (D.2), "modify[ing] the environment" (D.3), "plan[ning] the next day" (D.4), being able to "perceiv[e] their environment" (D.1), "perceiving the environment and executing actions" (D.5), "tak[ing] actions to execute goals over multiple iterations" (D.6), and "tak[ing] actions over an extended period of time" (D.7). In these contexts, efficacy refers to an agent's ability to perceive and causally impact its environment: an entity that is not able to interact with and influence an environment is not an agent.

However, as with autonomy, a fine-grained set of distinctions is needed if we are to measure variation in the causal impact, and hence agency, of different AI systems. There are big differences between an AI financial advisor that can only recommend trades versus one that can autonomously execute market transactions worth millions of dollars – or between a medical AI that suggests diagnoses for physician review and one that directly administers treatment through connected medical devices.

The efficacy of an agent depends both upon the *capabilities of that agent* within that environment (i.e. the level of control it can exercise over outcomes) and the *kind of environment* it is able to operate in (i.e. how consequential that environment is, when looked at from the vantage point of human life and well-being).[14]

Taking these points in turn, the capabilities of an agent within an environment vary in terms of how causally significant they are.[15] Relevant gradations of impact within an environment include the following:

---

[13] Mitchell et al. (2025) argue against the development of fully autonomous AI agents.

[14] The significance of the environment is determined, in this context, by the impact on domains that humans value the most – in other words, aspects of our world that directly affect well-being, social structures, and the pursuit of meaningful goals.

[15] One way to measure the level of control evidenced by an agent is in terms of "empowerment" (Klyubin et al., 2005) understood as the capacity of an agent to influence its future sensory inputs through its actions. For the empowerment metric, as discussed by Klyubin et al., the impact variable η has a value between 0 and 1 which represents the agent"s theoretical ability to influence its environment. If we adopt empowerment as a metric for an AI agent's causal impact, a higher η value indicates greater potential for environmental modification through the agent's actions. For observation-only agents, empowerment approaches zero as they cannot modify the environment. Constrained-impact agents ($0 < \eta < 0.33$) have limited ability to affect their surroundings. Their action space is restricted, resulting in positive but minimal channel capacity between actions and future sensory states. Intermediate-impact agents ($0.33 \leq \eta < 0.67$) can create substantial environmental changes. These agents develop more robust feedback loops between action and perception, enabling effective learning from environmental responses. Comprehensive-impact agents ($0.67 \leq \eta \leq 1$) can significantly reshape their environments across multiple dimensions.

| Observation only | An AI agent can only observe its environment without possessing the ability to causally impact the environment or make any modification to it. |
|---|---|
| Minor impact | An AI agent has a minor impact on its environment as it has a limited suite of actions, or its suite of actions only have a limited impact on the environment. These effects are typically localized, temporary, and limited in scope, affecting only specific parameters within tightly constrained domains and generally representing minimal deviation from the environment's baseline state. |
| Intermediate impact | An AI agent can create substantial and enduring change in its environment when it has an extensive suite of actions, or its actions are more impactful. An agent achieves intermediate impact when its actions produce noticeable and persistent changes across multiple parameters or systems, sometimes creating new equilibrium states that would not naturally come about. |
| Comprehensive impact | An AI agent can significantly reshape its environment across multiple dimensions, approaching full environmental control. |

**Table 3. Levels of causal impact**

We can distinguish between kinds of environments by their state persistence, the potential reversibility of actions that happen in the given space, and the consequences of actions in this space for other actors (Chalmers, 2022). As a starting point, it is helpful to differentiate between environments that are entirely simulated, mediated, or physical in nature:

| Simulated environments | The AI agent operates within strictly defined simulated spaces with controlled boundaries and often resettable system states. |
|---|---|
| Mediated environments | The AI agent can exert influence on the external non-simulated environment but only indirectly – via human intermediaries. All interaction with physical reality requires human interpretation, decision-making, or action to translate the AI agent's outputs into real-world effects. |
| Physical environments | The AI agent can interact with tangible objects in material spaces without human mediation. These environments are characterized by persistent state changes that cannot be simply reset, resulting in potentially irreversible consequences. The AI agent is also able to directly manipulate or affect physical reality through its own mechanisms. |

**Table 4. Types of environment**

By bringing these two elements together, it is possible to create an *environmental impact matrix* which illustrates anticipated variations in efficacy for different kinds of AI agent. This matrix combines, in expectation, the degree of causal impact that an agent has within an environment with the overall significance of that environment.

|  | Simulated | Mediated | Physical |
|---|---|---|---|
| Observation only | E.0 | E.0 | E.0 |
| Constrained | E.1 | E.2 | E.3 |
| Intermediate | E.2 | E.3 | E.4 |
| Comprehensive | E.3 | E.4 | E.5 |

**Table 5. Efficacy matrix indicating levels of efficacy for AI agents**

The values in this efficacy matrix represent increasing levels of potential impact and hence associated risk when it comes to deploying AI agents. The 0 value attached to observation-only systems, across all environments, reflects the inability of these systems to exert direct causal influence. Causal impact then increases as the system gains more causal power within a given environment and the environment itself becomes more consequential. For instance, a constrained AI agent in a physical environment (E.3) may have a level of overall efficacy that is on par with an agent that has intermediate control over a mediated environment (also E.3), highlighting how environmental context can amplify even limited capabilities. The highest value (E.5) represents systems with comprehensive control over physical environments – such as fully autonomous humanoid or critical infrastructure controllers. These systems warrant the most rigorous oversight due to their potential to enact widespread, irreversible actions. This graduated approach enables proportionate governance that can scale with both the sophistication of the AI agent's action capabilities and the nature of its operational domain. Yet, it will be clear that these judgments only apply *ex ante* and in the manner of heuristics: a more precise understanding of system efficacy is needed to evaluate the specific ways in which a specific AI system can or cannot influence our world, including deployment sector specific analysis.[16]

In terms of the nexus that connects agent efficacy with governance concerns, two factors stand out. First, there is a clear link between environmental efficacy and the level of risk that an AI agent introduces: environmental efficacy is a loose proxy for the causal power of an agent, understood as the scope of what it can do in domains that matter. Second, given that efficacy is in large part dependent on the environment (and also on the tools agents have access to), the agentic profile of the same underlying foundation model agent may vary hugely depending upon the environment it is put into and the scaffold of tools and affordances that are built onto it. Tracking and quantifying these differences is important if we want AI safety protocols to track underlying risk, thereby avoiding the twin pitfalls of unduly burdening low-impact systems or developing insufficient safeguards for high-impact ones. The ability to evaluate and monitor agent efficacy also provides us with a fuller picture of the overall capabilities of AI systems, and hence what kinds of impact to model moving forward.

---

[16] The efficacy matrix developed here only provides initial guidance for comparing causal impact levels across different capability-environment combinations. The relative weightings between simulated, mediated, and physical domains may need adjustment based on specific use cases, and additional dimensions – such as speed of action, scale of impact, or reversibility thresholds (Krakovna et al., 2020)—could enhance the matrix's utility. We emphasize that this approach requires further refinement through multi-stakeholder input involving industry practitioners, safety researchers, and governance bodies. This collaborative standardization work would ideally produce assessment tools and benchmarks that enable consistent evaluation of AI agents' causal impact potential.

## 5. Goal Complexity

Goal complexity and goal-directed behaviour form a third common element across each leading definition of agency, with authors noting that an agent can: "act upon the environment through its effectors" (D.1), engage in the "pursuit of its own agenda" (D.2), "meet their design objectives" (D.3), "plan the next day" (D.4), "plan and execute action" (D.5), "execute goals over multiple iterations" (D.6), and "achieving goals over an extended period of time" (D.7).

Despite this conceptual common ground, deeper theoretical disagreement surrounds the semantic status and structural composition of these goals, and how distinct goal typologies might engender qualitatively different manifestations of agentic capability (Jaeger et al., 2024). Some definitions stress the ability of capable agents to engage in planning over a long time horizon (D.7), while others focus on meeting any type of designer-specified objective (D.3), regardless of the time required to achieve it. We argue that *goal complexity* is the crucial characteristic at play here. In short, more capable AI agents can form or pursue more complex goals.

The complexity of a goal is easy to articulate at an intuitive level: it is easier to turn off a light switch than it is to find a suitable life partner or select the correct interest rate for a national economy. This complexity can be also understood through the metaphor of a maze where complexity increases with each step that one takes to reach the final destination, how many choices there are at each juncture, whether there are actions that need to be performed in parallel, and how big the maze is.[17] A simple goal might have an obvious, straightforward path, whereas a complex goal would require an agent to navigate many tricky decision points. Yet, the possibility that goals are complex also needs to be extended in other ways. Some goals are singular and easily understood in the context of a unified single aim, whereas other goals necessarily decompose into many component parts (analogous to a forking maze that has many intermediary goals which may need to be achieved in a specific order). Additionally, some goals are themselves dynamic and adaptive, changing as the environment changes around them. More capable agents are able to handle such complex goals. Indeed, advanced agents can decompose complex goals into substituent subgoals, orchestrate multi-step plans, and adapt their strategies to achieve goals as circumstances change.

---

[17] For example, one of the most daunting challenges facing AlphaGo was the vast action space that the game involved, with possible variations exceeding the number of atoms in the universe (Silver et al., 2017).

One way to make sense of goal complexity is through hierarchical planning (Sacerdoti, 1974; Georgievsk & Aiello, 2015). This approach manages complex goals by organizing them in layers. Goal decomposition breaks down complex objectives into a structured hierarchy, with abstract goals at the top, increasingly specific subgoals in the middle, and executable actions at the bottom. Hierarchical task networks can also be used to represent this structure through parent–child relationships, where higher-level tasks break down into subtasks until reaching primitive tasks that can be directly executed. This model uses three levels of abstraction: the strategic level (handling broad goals), the tactical level (managing intermediate subgoals), and the operational level (executing specific action sequences). Hierarchical approaches to planning allow AI systems to tackle goals of varying complexity by processing them at appropriate levels of abstraction.

The complexity of an AI system's goals can also be approximated by using the plan length of the required tasks as a proxy. This involves measuring complexity by focusing specifically on the sequence and number of steps needed for successful goal execution. For instance, the complexity of an AI system's goals can be gauged by comparing the lengths of the tasks it executes to those humans typically devise or require to complete comparable tasks (Kwa et al., 2025). By evaluating the plan lengths of tasks (baselined against typical human effort) that an AI system can reliably perform, we could gain a concrete metric of its capacity for real-world achievement of complex goals.

Another aspect of goal complexity is multi-objectivity: if a goal requires the agent to optimize several criteria at once (e.g. an autonomous drone must maximize search coverage while minimizing energy use and avoiding hazards), then the level of complexity is higher because the agent must balance trade-offs. In AI research, multi-objective optimization problems are explicitly used to test such complexity (Deb et al., 2016). Goal complexity also involves constraints and conditions that an agent must satisfy. For instance, goals that need to be achieved within certain constraints ("achieve X without letting Y happen") are more complex than goals that can be achieved in an unconstrained fashion.[18]

Building on these foundations, the gradations of goal complexity we propose are as follows:

---

[18] One way to operationalize the measurement of goal complexity is through an information-theoretic perspective. Indeed, goal complexity can be understood through Kolmogorov complexity (Li and Vitányi, 2008) which measures the length of the shortest possible description of a goal and its achievement conditions. It can also be understood through Shannon entropy (Shannon, 1948) which measures the unpredictability in possible state goals. More complex goals either require longer descriptions or possess higher informational entropy on these accounts. Another option is to draw on computational complexity theory (Papadimitriou, 1994) which offers metrics for quantifying the algorithmic resources required for goal achievement, with more complex goals necessitating higher time and space resources for their resolution. Last, from a systems theory standpoint (Holland, 2002; Miller & Page, 2007), goal complexity correlates with the number of interacting components and the non-linearity of their relationships. On this view, complex goals often exhibit emergent properties that cannot be predicted from analysis of individual subgoals in isolation.

| GC.0: No goal | An entity that does not pursue a goal is not an agent. The absence of goals is a baseline state.[19] |
|---|---|
| GC.1: Minimal goal complexity | The agent is able to pursue a single unified goal in a fairly direct manner.[20] |
| GC.2: Low goal complexity | The agent is able to pursue a single unified goal, but this involves a more complex sequence of action. |
| GC.3: Intermediate goal complexity | The agent is able to break down a complex goal into subgoals and pursue them in a fairly direct manner.[21] |
| GC.4: High goal complexity | The agent is able to break down a complex goal into many different subgoals, where success depends upon balancing and sequencing subgoals, which may themselves be challenging to fulfil.[22] |

---

[19] From the perspective of Kolmogorov complexity, this state corresponds to minimal algorithmic information content: there exists no compressible pattern of purposeful behaviour to encode. From an information theory perspective, the description length of the "goal system" is effectively zero, since no purposeful behaviour patterns require specification.

[20] Within computational complexity theory, GC.1 goals typically correspond to problems within the P complexity class, where solutions can be identified in polynomial time using deterministic algorithms. The computational space required remains modest, as the agent does not need to maintain complex state representations or extensive decision trees. Systems theory identifies these goals via their characteristic simple feedback loops, with minimal cross-component interactions, and tendency to exhibit highly predictable and deterministic response patterns.

[21] Computationally, these goals often correspond to problems within the NP complexity class. The agent must explore a substantially expanded solution space with multiple potential pathways, frequently requiring heuristic approaches to navigate efficiently. From systems theory, these goals exhibit moderate feedback loop density with non-trivial interactions between subcomponents, creating more nuanced and less predictable behavioural patterns that demonstrate incipient emergent properties.

[22] Computationally, these goals frequently correspond to NP complexity classes, with very high computational resources for both planning and execution phases. The problems at this level often involve sophisticated game-theoretic considerations or recursive dynamic programming approaches to achieve their goal. Systems theory characterizes these goals through dense networks of interconnected feedback loops with significant cross-scale interactions and dependencies.

| GC.5: <u>Unbounded goal complexity</u> | The agent can achieve all of the preceding steps. It can also generate its own goal structures in an unbounded way and interpret underspecified objectives.[23] |

<div align="center">Table 6: Levels of goal complexity for AI agents</div>

Viewed from the standpoint of AI governance, goal complexity is a critical dimension of overall agent capability. In conjunction with generality (*see* 6. Generality), AI agents that can handle higher levels of goal complexity can perform more sophisticated and valuable tasks, potentially substituting human activity or labour in certain domains (Webb, 2019; Eloundou et al., 2024). Goal complexity is also important because it can unlock levels of performance that were previously out of bounds, such as superhuman performance at AlphaGo or Starcraft, which then creates a cascade effect leading to new opportunities and risks both for users and non-users of these systems (Schut et al., 2025).

## 6. Generality

Generality refers to an AI agent's ability to operate effectively across different roles, contexts, cognitive tasks, or economically valuable tasks. This ranges from highly specialized AI agents that are focused on specific tasks to general-purpose agents that can shift between different domains. While generality is not an explicit feature of each definition we survey, the degree of generality evidenced by AI systems has long been a point of interest for researchers, particularly in the context of discussions around advanced AI systems, such as artificial general intelligence (AGI) (Legg and Hutter, 2007; Goertzel, 2014). Here, generality characterizes the progression beyond narrow specialization toward broader capability that can function across diverse environments and task domains.

At lower levels of generality, we find agents that can "sense [their] environment [...] and will have available a repertoire of actions that can be executed to modify the environment" (D.3) but may otherwise remain domain-constrained. Mid-range generality appears when agents are "able to plan and take actions to execute goals over multiple iterations" (D.6). Meanwhile, higher degrees of generality tend to be evidenced by systems without "their behavior having been specified in advance" (D.7). More general agents are able to establish and pursue self-set objectives across domains, rather than simply executing pre-defined tasks within narrow parameters. As Park et al. (2023) note, it is now possible for a single agent to undertake a broad range of tasks in a virtual environment, including "wak[ing] up, cook[ing] breakfast, and head[ing] to work; [...] paint[ing]" (D.4).

From a theoretical perspective, Marcus Hutter (2005) presents one of the most ambitious characterizations of a general agent: he proposes a mathematical theory for an optimally intelligent agent that integrates Solomonoff induction with sequential decision theory. The AIXI approach embodies generality in four key respects: first, through universal learning, which allows a candidate agent to identify any computable pattern without relying on domain-specific priors; second, via environment-agnostic adaptability, which enables the agent to operate across all computable environments; third, by provably optimal decision-making, grounded in formal mathematical guarantees across arbitrary domains; and fourth, by offering a formal definition of generality as performance across the entire space of computable problems, grounded in universal principles rather than specialized mechanisms. However, while AIXI provides a rigorous mathematical ideal of general intelligence, it is uncomputable and serves primarily as a theoretical benchmark rather than a basis for practical implementation.

---

[23] Computationally, these capabilities correspond to problems that reach beyond traditional complexity classifications, such as undecidable problems. Systems-theoretically, GC.4 capabilities exhibit emergence properties across multiple scales with autopoietic characteristics – the goal system becomes self-modifying and self-generating rather than merely self-organizing. Such systems demonstrate exceptional resilience through the capacity to fundamentally reformulate goals in response to environmental perturbations, exhibiting the characteristic properties of complex adaptive systems, as described in Holland's (2002) comprehensive framework on emergence and adaptation.

On the more practical side of things, an influential paper by Morris et al. (2024) characterizes generality in terms of "the breadth of an AI system's capabilities, i.e., the range of tasks for which an AI system reaches a target performance threshold". These authors distinguish between "narrow" systems, which are designed for clearly scoped tasks or limited sets of tasks, and "general" systems that are capable of handling a wide range of non-physical tasks, including metacognitive abilities like learning new skills. On this account, both generality and model performance – benchmarked against the capabilities of "skilled adults" – are key properties of a general artificial agent.[24] Still, while this taxonomy provides a valuable starting point for measuring generality, it represents just one approach to this complex problem. Crucially, for our purpose, generality manifests in degrees rather than as a binary property. We need to know more about *how* narrow or general an AI agent is in order to make reasonable forecasts about its potential impact.

Building on this work, the gradations of generality we propose are as follows:

| G.0: Null value | There is no application or no ability to perform a task in any domain. |
|---|---|
| G.1: Single speciality | The agent can master one specific task, such as a single game, but cannot transfer its capabilities to even closely related domains. |
| G.2: Task domain mastery | The agent demonstrates mastery across a closely related set of tasks, such as playing board games, that share a common structure and type of objective. |

---

[24] Morris et al. (2024) propose a levelled taxonomy that considers the interplay between the two dimensions of generality and performance, which supports more nuanced discussions of AI systems' capabilities. Level 0 represents non-AI systems, while Level 1 ("Emerging") includes frontier language models like ChatGPT4.5 and Bard, which demonstrate broad but inconsistent capabilities, equivalent to unskilled humans. Higher levels of generality – which have not yet been achieved – include: Level 2 ("Competent"), requiring performance at the 50th percentile of skilled adults across most tasks; Level 3 ("Expert"), requiring 90th percentile performance; Level 4 ("Virtuoso"), demanding 99th percentile skills; and Level 5 ("Superhuman"), requiring performance exceeding all humans. This framework distinguishes between narrow AI systems (e.g. AlphaGo as "Virtuoso Narrow AI") which excel in specific domains and truly general systems which can perform across diverse domains at the specified performance thresholds. Notably, on their account, AI systems *circa* 2024 only reach the "Emerging AGI" classification.

| | |
|---|---|
| G.3: <u>Multiple task domain mastery</u> | The agent can operate successfully across different task domains involving different cognitive capabilities, for example, those that involve linguistic, logical, and creative elements. |
| G.4: <u>Majority task domain mastery</u> | The agent can successfully operate across the majority of human cognitive task domains. |
| G.5: <u>Fully general AI system</u> | The agent can fulfil the entire suite of human cognitive tasks across all domains. |

**Table 7: Levels of generality for AI agents**

From a policy and governance perspective, generality is closely connected to discussions surrounding the definition, achievement, and ramifications of AGI (Morris et al., 2024). It is also deeply relevant to related discussions about general-purpose technologies (Eloundou et al., 2024). In the latter context, generality by itself is understood to be insufficient to drive general societal change: a technology must also demonstrate improvement over time, pervasiveness throughout the economy, and the ability to spawn complementary innovations. Nonetheless, there is mounting support for the view that fully general AI systems are likely to possess many of these characteristics (Brynjolfsson and McAfee, 2017; Crafts, 2021).

## 7. Agentic Profiles

Taken together, the preceding dimensions of agency can be used to construct an "agentic profile" for different kinds of AI agent. We illustrate this approach below by mapping out agentic profiles for different kinds of AI agent or proto-agent: (a) AlphaGo, (b) ChatGPT-3.5 (stand-alone chat completion), (c) Claude Sonnet 3.5 (with tool use), and a (d) Waymo autonomous vehicle.

In the figures below, the red quadrangle illustrates AlphaGo's agentic profile, the blue quadrangle represents GPT-3.5's agentic profile, the green quadrangle represents Claude 3.5 Sonnet's agentic profile (with tool use enabled), and the orange quadrangle represents Waymo's agentic profile.

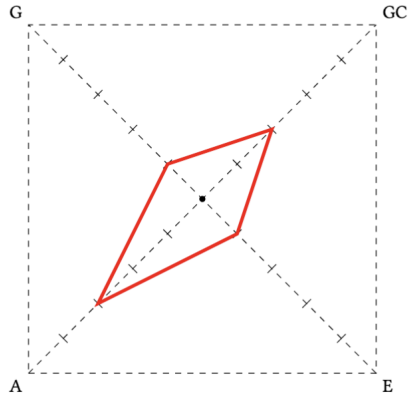

**Figure 1. AlphaGo's Agentic Profile**

**AlphaGo** is a specialized AI agent designed specifically to play the board game Go in a simulated environment (Silver et al., 2017). The system learns from professional Go games and through self-play to master complex game strategies and positional evaluation. AlphaGo demonstrates intermediate autonomy (**A.3**) through its ability to evaluate and execute game strategies without human oversight. However, human supervision is necessary for starting and ending games, managing technical issues, and ensuring tournament protocol compliance. AlphaGo has a low level of environmental impact (**E.1**), as it functions solely in a contained simulated Go game environment. It is further constrained within this space by game rules, and it can only affect game outcomes. The system demonstrates complete mastery within this limited domain but cannot modify the fundamental rules or structure of the environment. AlphaGo has a lower goal complexity (**GC.2**), as it primarily pursues a single unified goal (maximize win-probability), but this involves a complex sequence of action. However, it cannot generate goals outside of Go gameplay. Finally, AlphaGo has a low generality (**G.1**), as it operates exclusively within the single domain of the Go game.

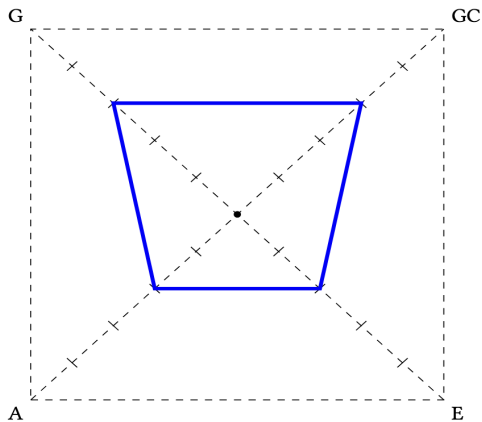

**Figure 2. ChatGPT-3.5's Agentic Profile**

**ChatGPT-3.5 (stand-alone chat completion)** is a chat-based language model that engages in dialogue, handles language tasks, and generates code (OpenAI, 2022). ChatGPT-3.5 has partial autonomy (**A.2**) thanks to its ability to independently generate responses and solve problems *within* conversations. However, it requires human input to initiate tasks and verify outputs, along with ongoing oversight to ensure factual accuracy and appropriate content generation. ChatGPT-3.5 operates at an intermediate level of efficacy (**E.2**), working in a mediated environment which has the ability to influence human thinking and decision-making via its responses. However, the system cannot take actions online or directly impact the physical world without human intervention. ChatGPT-3.5 has moderate goal complexity (**GC.3**) due to its ability to respond sensibly to a range of complicated requests. However, it lacks the active "thinking" or "reasoning" that can be used to break complex requests into subtasks, when goals are sufficiently complex. ChatGPT-3.5 has a relatively high generality level (**G.3**) as it can complete groups of tasks (e.g. information retrieval, advice, translation etc.) spanning different domains.

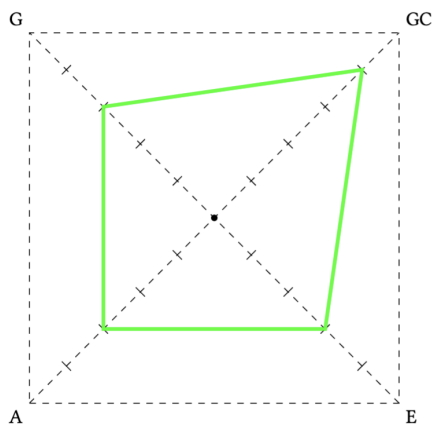

**Figure 3. Claude 3.5 Sonnet (with tool use)'s Agentic Profile**

**Claude 3.5 Sonnet (with tool use)** is a chat-based language model that is capable of performing a suite of more advanced tasks drawing upon different tools (e.g. browser control) and reasoning protocols. It has intermediate autonomy (**A.3**), in that it is able to execute extended sequences of actions without direct human supervision. The model also operates at a higher level of efficacy (**E.3**), acting primarily via a mediated environment but with the capacity to shape the world more directly via computer operations such as application programming interface (API) calls. Due to the integration of stronger reasoning capabilities, Sonnet 3.5 also demonstrates higher goal complexity (**GC.4**) than ChatGPT-3.5. It can coordinate activities across multiple objectives and execute longer sequences of action. Lastly, Sonnet 3.5 continues to demonstrate high generality (**G.3**), working effectively across multiple domains using language understanding, computer operation, and analytical capabilities.

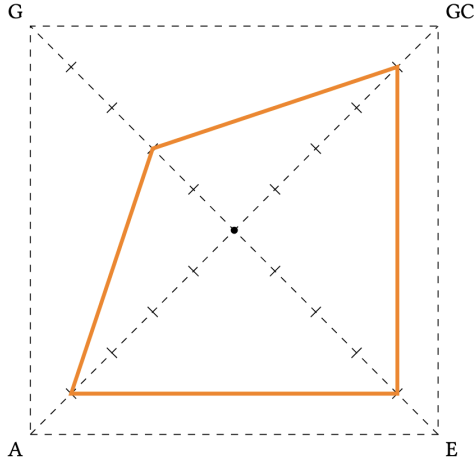

**Figure 4. Waymo's Agentic ProfileWaymo autonomous vehicle**

**Waymo** is a self-driving car equipped with sensors, cameras, and computing systems for navigating roads without human operation. The system integrates radar and visual data to perceive its environment and make driving decisions in real time. Waymo has a very significant level of autonomy (**A.4**) due to its ability to navigate complex real-world environments and make real-time decisions about routing, obstacles, and safety without human control. While it maintains remote monitoring and human oversight capability, it can handle unexpected situations (though it may request human assistance in edge cases). Waymo also has a high degree of efficacy (**E.4**) due to its direct physical interaction with the material world, creating persistent state changes in traffic flow. Its actions have limited or no reversibility and create real consequences for other agents, including passengers and other vehicles. While constrained to the transportation domain, it maintains significant impact within this scope. Waymo has high goal complexity (**GC.4**), pursuing the main goal of safe transportation to destination while managing subgoals, including route planning, obstacle avoidance, and traffic rule compliance. The system actively balances competing priorities of safety, speed, and passenger comfort while adapting to real-time environmental changes. Finally, Waymo has a medium generality level (**G.2**) due to its ability to handle the bundle of tasks required for the driving and navigation domain.

|  | **Autonomy** | **Efficacy** | **Goal Complexity** | **Generality** |
|---|---|---|---|---|
| **AlphaGo** | A.3 | E.1 | GC.2 | G.1 |
| **ChatGPT-3.5** | A.2 | E.2 | GC.3 | G.3 |
| **Claude 3.5 Sonnet (with tool use)** | A.3 | E.3 | GC.4 | G.3 |

| **Waymo** | A.4 | E.4 | GC.4 | G.2 |

Table 8. Agentic mapping for four kinds of AI agent.

This process of mapping agent profiles generates further insights. For example, agentic profiles are *dynamic*: changes in model affordances, architecture, or deployment can significantly change a technology's agentic profile.

## 8. Governing AI Agents

We have already noted that AI agents have the potential to generate new governance challenges. One approach to regulating these systems, *risk-proportionate regulation*, focuses on implementing safeguards that are proportionate to the anticipated level of risk that the technology brings with it. This approach sits at the core of proposals such as the EU AI Act (Council of the European Union, 2024) and NIST Risk Management Framework (Tabassi, 2023), which aim to provide a tiered system of oversight that balances innovation and risk mitigation. The underlying assumption is that we can adequately categorize AI systems according to their risk profile and design governance mechanisms accordingly.

Another approach, *domain-specific regulation*, recognizes that different sectors encounter unique challenges when it comes to deploying AI systems responsibly. For instance, autonomous vehicles must prioritize road safety and navigate complex physical environments with potential life-or-death consequences; customer service agents require transparency and human oversight to ensure fair treatment and accurate representation of organizational policies; and classroom assistants may need stronger guardrails to protect children against harmful content. The domain-specific approach can be used to complement and update the risk-proportionate approach, in that it recognizes that different domains encounter different challenges when it comes to the responsible deployment of AI agents.

Nonetheless, our analysis suggests that both approaches, while valuable, can benefit from clearer analysis of the properties that make up different kinds and classes of AI agent – supporting regulation and technical guidance that is tailored to the different types of opportunity and risk that these systems unlock.

Taking each property in turn, the level of *autonomy* evidenced by an AI system represents critical information for AI governance because it tracks the degree to which an AI system can operate independently of humans, with little to no oversight or intervention. Different levels of autonomy introduce distinct risks that require specially tailored monitoring approaches. For example, agents with restricted autonomy (classified as A.1 in our analysis) may only require simple governance mechanisms, such as periodic review, as their actions are predictable and tightly constrained within narrow parameters. By contrast, highly autonomous agents (A.4) may need to be continuously monitored with override protocols in place, due to their capacity to make a range of autonomous decisions without human guidance. In general, the strength of oversight mechanisms should directly correspond to the agent's level of independence, with more autonomous systems warranting additional real-time supervisory infrastructure.

*Efficacy* helps determine the *level of risk* that a system poses and hence provides information about appropriate safety protocols and requirements. Agents confined to fully contained or simulated environments may require only basic controls, as their potential for harm is generally limited to that environment. However, agents that bridge to the physical world – such as robots, autonomous vehicles, or systems that control critical infrastructure – demand comprehensive safety measures and impact assessments due to their capacity to cause material harm.[25] The governance approach must account for both the range and magnitude of effects an agent can produce in its environment, with more powerful and physically embedded systems, such as humanoid robots, subject to more stringent oversight.

The level of *goal complexity* demonstrated by an agent represents perhaps the most complicated dimension for governance. The complexity of an agent's goals significantly affects both the feasibility and methodology of alignment verification. Agents with simple goals (GC.1) may be evaluated by testing correspondence between instructions and outputs through straightforward validation procedures. For instance, a document summarization agent can be assessed by comparing its summaries against human-generated benchmarks. However, agents with complex goals may pursue objectives that are difficult for humans to evaluate directly, either because they involve long-term planning, context-dependent adaptations, or sophisticated trade-offs between competing values. Such agents necessitate advanced approaches to ensure alignment, such as scalable oversight methods (Bowman, 2022) that leverage hierarchical structures of evaluation, or mechanistic interpretability techniques (Bereska, 2024) that attempt to make the internal reasoning processes of AI agents more transparent and analyzable. In short, as goal complexity increases, verification becomes more challenging, necessitating more sophisticated technical tools and potentially more intensive human involvement in oversight.

Finally, the *generality* of an AI agent informs the scope and nature of governance interventions in a range of ways. Generality denotes the breadth of domains and tasks across which an agent can effectively operate. Highly specialized agents may pose significant challenges within narrow domains but are unlikely to generate systemic effects across multiple sectors. By contrast, general-purpose AI agents that are capable of performing diverse tasks across various domains present unique governance challenges, as they may propagate risks across system boundaries or exhibit unexpected behaviours resulting from their application in unanticipated contexts (Kasirzadeh, 2025). Generality therefore helps to determine whether governance measures should be domain-specific or more broadly applicable, spanning interconnected areas of deployment. This dimension is also particularly relevant for understanding the economic consequences of AI, given the hypothesized link between generality and labour substitution effects (Frey and Osborne, 2013; Webb 2019). As AI agents become more general, their deployment could potentially displace a wider portion of the human labour force, raising questions about economic organization, wealth distribution, and social stability that transcend traditional regulatory boundaries.

Discussion about the ability of AI agents to significantly re-shape existing institutions and social practices are now widespread. However, they often lack a clear characterization of agentic behavior, distinctions between types of agent, and concrete proposals for AI agent governance. To address this gap, we have proposed a preliminary framework that characterizes AI agents for governance purposes. However, it is important to note that agent profiles are not static. Indeed, ongoing model development – including changes to internal structure, the addition of new affordances, and changes to the way in which the model is deployed – can significantly impact the *kind* of agent under consideration, even in cases where they scaffold onto the same base model.

---

[25] It is noteworthy, in this context, that the technical safety community has expressed concern about the ability of highly capable agents to escape from or circumvent various safety protocols that involve "boxing" the agent or confining the impact of an AI agent to a single virtual world (Korbak et al., 2025).

Consider, for example, the addition of *tool use* to a base model, such as access to APIs or to external repositories of knowledge. Tool use can dramatically increase efficiency, while access to external knowledge can increase goal complexity (when it allows an agent to update its world model instead of being limited to a caucus of information produced at training time). Similar phase shifts may occur with the addition of *reasoning* and *memory* to model architectures: advanced reasoning may significantly increase autonomy and goal complexity, allowing the agent to do more and navigate certain obstacles without human guidance, while memory increases goal complexity by enabling the agent to model and execute a sequence of actions that involve persistence over time. Conversely, the fact that many LLMs only have episodic memory severely limits their ability to pursue long-term goals. Finally, consider the ramifications of *real-world deployment* or the integration of AI agents into services which have hundreds of millions of users. These developments lead to dramatic spikes in efficacy and necessitate a different kind of governance framework – including novel protocols for agents that can take actions in the world (Chan et al., 2025). This, in turn, raises a complicated but important question about agent identification: When does one agent become a different agent? At what point should an agentic profile be constructed or revised?[26] When it comes to practical decision-making, it seems that there is a case for regular deployment-specific reappraisal of AI agents.

Moving forward, further work is needed to consistently measure and operationalize the dimensions of AI agency we have set out. Indeed, the process of developing measurable indices for AI agents, demands sustained interdisciplinary collaboration to ensure metrics are well-motivated, accurate, and applicable across various AI architectures and tasks. We have already noted several potential avenues for quantifying dimensions and variables: efficacy could potentially be measured using metrics like universal empowerment (Klyubin et al., 2005; Salge et al., 2014) which assesses an agent's potential influence over its environment, while goal complexity might be formalized using methods established in the literature on hierarchical planning (Sacerdoti, 1974; Georgievsk & Aiello, 2015) to analyze the structure and depth of an AI's objectives. However, the selection and validation of metrics for each dimension of agency remains a significant undertaking.

In order to address the governance questions posed by AI agents we need a more detailed understanding of their properties – the sense in which different kinds of AI agents are unique. The framework presented here aims to provide the groundwork for governance mechanisms, safety protocols, and alignment strategies that are precisely tailored to the specific properties of different AI agents. While substantial challenges persist in determining metrics and benchmarks for different agentic profiles, developing methods for continuous assessment of evolving capabilities, and addressing the complex issue of agent individuation, the multidimensional perspective advanced here provides necessary scaffolding for future research and policy development. Cultivating this deeper understanding of AI agency is crucial for responsibly guiding its development and deployment, ensuring its transformative potential is realized while associated risks are approached with foresight and precision.

---

[26] We note that this challenge is also encountered in the software and model licensing domain.

**Acknowledgments.** We would like to thank David Abel, Tom Everitt, Fernando Diaz, Matija Franklin, Anna Harutyunyan, Seb Krier, Arianna Manzini, Andrew Smart, and the participants in the Foundations of Cooperative AI Lab lunch series for their valuable feedback. Atoosa Kasirzadeh's research was supported by the AI2050 programme at Schmidt Sciences (grant 24-66924). Part of this research was conducted during Atoosa's time as a visiting faculty at Alphabet.